\documentstyle[preprint,aps]{revtex}
\begin{document}
\draft
\preprint{Lecce University}
\title{The Super P-Brane Scan and S Duality}
\author{ 
Khaled Abdel-Khalek\footnote{e-mail: khaled@le.infn.it}} 
\address{Dipartimento di Fisica - Universit\`a di Lecce\\
- Lecce, 73100, Italy -}
\date{July, 1996}
\maketitle
\begin{abstract}
Taking into account the recent dualities we rederive
the super p-brane scan. Our main results are the importance of the 
metric's signature and the existence of an S  
self-dual super 5-brane at D=14 with signature (7,7) or (11,3). 
\end{abstract}

\widetext

Recently, new developments in superstring theory have led to 
major advances in our understanding of the non-perturbative 
regime of the theory. Dualities taught us that there exists 
an underlying unique theory, it may allow different vacua. 
We have mainly two types of dualities S and T, sometimes, 
they can unify into U duality. It can happen that theory A 
at strong coupling is equivalent to theory B at weak 
coupling, they are said to be S dual, while, if theory A is 
compactified on a space of volume  $\tau$ is equivalent to 
theory B compactified on a space of volume $\sim 1/\tau$, then 
they are called T dual.
The S duality can be considered as a generalization of the 
Manton-Olive electromagnetic duality 
\cite{r1} where a theory at small 
coupling is mapped to its solitonic partner at strong 
coupling.

One of the striking consequences of these new achievements 
is the existence of the eleven dimensions 
M-theory\cite{r2}, a quantizable version of 
the supermembrane and another more fundamental the  F-theory
\cite{r3}.
The accumulating evidences of this F-theory indicate that 
it lives in 12 dimensions of signature (10,2). At first 
sight, it seems strange to have a physical theory in 
$D > 11$, since, a spinor in D dimensions has 
$2^{(D+2)/2}$ real components i.e in 12 dimensions,
the spinor will have 128 real components whereas any
physical theory (that does not couple to particles with 
helicity higher than two) have maximally 32 real components.
Actually, this statement overlooked 
something very important since it ignores the 
signature of the underlying space-time. 
Supersymmetric theories are very sensitive to the 
dimension of the spinor,  the nature of the spinorial 
representation of the lorentz group (being real 
or imaginary) which depends on the dimension and the 
signature of space-time and moreover the possible
conditions that one can impose on his spinor and this is 
exactly what we are going to investigate in this note.

The construction of any supersymmetric theories is 
systematic and straightforward once the automorphism
group of its corresponding Clifford algebra is given
\cite{r7}. So the knowledge of Clifford algebra is vital. 
Such algebra are completely classified by mathematicians for 
any generic D dimensions  with signature (s,t)\cite{r4,r5}. I have 
reproduced it in 
Table (\ref{table1}) as we will use it heavily.
\vspace{3cm}
\begin{center}{- TABLE I -}\end{center}
\vspace{3cm}

Using Table(\ref{table1}) we can
generalize the results of \cite{r7} to any dimension D with 
signature (s,t) we  have the following properties for our 
spinor 
\begin{eqnarray}
s - t &=&~~2,~6~~~~~\mbox{mod}~~8~~~~~\mbox{complex,}\\
      &=&~~0,~1,~7~~\mbox{mod}~~8~~~~~\mbox{real,}\\
      &=&~~3,~4,~5~~\mbox{mod}~~8~~~~~\mbox{pseudoreal.}
\end{eqnarray} 
One can read easily the dimension of the spinor  from Table
(\ref{table1}). Their automorphism group is given by
\begin{eqnarray}
s - t &=&~~2,~6~~~~~\mbox{mod}~~8~~~~~\mbox{SU(N)$\times$U(1),}\\
      &=&~~1,~7~~~~~\mbox{mod}~~8~~~~~\mbox{SO(N),}\\
      &=&~~0~~~~~~~~\mbox{mod}~~8~~~~~\mbox{SO(N)$_+\times$SO(N)$_-$,}\\
      &=&~~3,~5~~~~~\mbox{mod}~~8~~~~~\mbox{USp(N),}\\
      &=&~~4~~~~~~~~\mbox{mod}~~8~~~~~\mbox{USp(N)$_+\times$USp(N)$_-$}.
\end{eqnarray}

In this paper, we will concentrate on the Weyl-Majorana cases 
as they are the most promising ones in higher dimensions.
It is well known 
that if we have a Weyl-Majorana on-shell (WM) spinor, it will reduce the 
number of components to the quarter.  
A Weyl-Majorana spinor is possible for D dimensions space-time of 
signature (s,t) if we can impose a Weyl condition\cite{r6}
\begin{eqnarray}
s-t =~~0~~~~~~~~~~~\mbox{mod}~~4,
\end{eqnarray}
while a Majorana condition is available for\cite{r6}
\begin{equation}
s - t =~~0,~1,~2~~~\mbox{mod}~~8.
\end{equation}
By combining the two conditions together, we have
\begin{equation}\label{mw}
s - t = 0 \ \mbox{mod}\ 8 ,
\end{equation}
another way to get this result is by combining a Majorana 
spinor
with a pseudo-Majorana.
One can check (\ref{mw}) explicitly by looking  in 
Table(\ref{table1}) and search for the $^2R$ cases. 
I have produced all the 
allowed physical cases in 
details in Table (\ref{table2}).
\vspace{3cm}
\begin{center}{- TABLE II -}\end{center}
\vspace{3cm}
Indeed in 12 dimensions a 32 real component  
spinor is possible for the following 
signatures (10,2) and (6,6). 
The (10,2) case has been noticed along time ago
in \cite{r6} and recently in \cite{r3,r8}.
Generally, a WM spinor is 
possible for any even 2n dimensions with signature (n,n). 
Such cases are very interesting as it is conjectured that 
their 
supersymmetric self dual theories are the generators 
of all known - and still unknown - possible integrable 
models\cite{r9,r10}. 
Fortunately for a Weyl-Majorana spinor, 
its automorphism group is 
well known and always of the type 
$SO(2^{(D-2)/2})_+ \times SO(2^{(D-2)/2})_-$.

The first application of Table(\ref{table2}) can be the 
scan for possible  (d-1)-superbranes \cite{r11}.
The standard analysis goes as the following : An important 
key in constructing these Green-Schwarz type  
superbranes is the existence of the $\kappa$ symmetry 
 which means that half of the spinor degrees of 
freedom are redundant and may be eliminated by a physical 
gauge choice. Let M be the number of real 
components of the minimal spinor and N the number of 
supersymmetries in D dimensions and let m and n be the 
corresponding quantities over the d worldvolume dimensions, 
the number of  bosonic effective degrees of freedom 
is
\begin{equation}
N_B = D - d
\end{equation} where we are working in the ``static gauge 
choice''. the number of on-shell fermionic degrees of freedom is
\begin{equation}
N_F = \frac{1}{2} m n = \frac{1}{4} M N.
\end{equation}
Worldvolume supersymmetry requires $N_B = N_F$, hence
\begin{equation}\label{ll}
D - d = \frac{1}{2} m n = \frac{1}{4} M N.
\end{equation}
After substituting for $m=2^{(d-2)/2}$ and $M=2^{(D-2)/2}$
we have
\begin{equation}\label{ll1}
D - d = \frac{1}{2} 2^{(d-2)/2} n = \frac{1}{4} 2^{(D-2)/2} N.
\end{equation}

We have shown the allowed values of p and n in 
the last column of Table (\ref{table2}). 
We note that $D_{max} = 12$ 
 whereas $d_{max} = 6$.
Hence, we have at max 
super 5-brane.
From table(\ref{table2}), we have the possible situations 
\begin{eqnarray}
d&=&2~~~~~D~=~3,~4,~6,~10.~~~~~~~~\mbox{string}\label{p1}\\
d&=&3~~~~~D~=~4,~5,~7,~11.~~~~~~~~\mbox{membrane}\label{p2}\\
d&=&4~~~~~D~=~5,~6,~8,~12.~~~~~~~~\mbox{3-brane}\label{p3}\\
d&=&5~~~~~D~=~6,~7,~9.~~~~~~~~~~~~~~\mbox{4-brane}\label{p4}\\
d&=&6~~~~~D~=~~~~8,~10.~~~~~~~~~~~~~\mbox{5-brane}\label{p5}
\end{eqnarray}
 
It is conjectured that all  super p-branes for $D<11$ are solitonic solutions.
But, for $D=11\ \mbox{and}\ 12$, they are 
the low energy limit of the M and F theory 
respectively.
From (\ref{p1}--\ref{p3}), we see three distinct series,
for any d=x, super (d-1)-brane exists for D=x+1, x+2, 
x+4, x+8, we will call
them the real, complex, quaternion and octonion series respectively.

Actually, we have  taken 
the signature of the p-brane worldvolume's metric to be that of
the minimal spinor,
if choose such metric to be (d-1,1) then m is minimal  only 
for d = 2,3,9,10,11 as can be shown directly from 
Table (\ref{table2}).
But this choice will affect only the number of supersymmetries
on the super p-brane world volume and will not make any further changes
to our results.

From the hierarchies of string/string, string/membrane ... dualities,
one would like to make the final theory distinguishable. A possible way
is to require it to be self dual with respect to the S duality.
From \cite{r11}, we find this 
possibility is only available for
\begin{equation}
D = 2(d+1),
\label{qw}\end{equation}
If we try to satisfy (\ref{qw}) for the real and complex series, d will
lead to  negative values of p whereas 
for
the other series,  
we have
\begin{eqnarray}
d&=&2\ \ D=~6~~~\mbox{the quaternion series}\\
d&=&6\ \ D=~14~~\mbox{the octonion series}
\end{eqnarray}
So, the final theory may be a super 5-brane in 14 dimensions 
but its existence is a puzzle. 
If we accept that the final theory is too
different then there should be a new ``generalized''
symmetry and there is no low energy limit's 
traditional super-gravitational
theory.
This result suggests to replace (\ref{p4}) and (\ref{p5})
by 
\begin{eqnarray}
d&=&5~~~~~D~=~6,~7,~9,~13.~~~~~~~~~~~~~~~~\mbox{4-brane}\label{p4a}\\
d&=&6~~~~~D~=~7,~8,~10,~14.~~~~~~~~~~~~~~\mbox{5-brane}\label{p5a}
\end{eqnarray}
In order to mach these results with
(\ref{ll}),
there should be a generalization of $\kappa$ symmetry,
for D = 7 and 14 super 5-brane (as the minimal spinor
in 14 dimensions with signature (7,7) or (11,3) has
64 real components)    
or we should include other fields so (\ref{ll}) will 
acquire new bosonic degrees of freedom as 
had been suggested in \cite{r13}. 
 
We can reach the conclusion that the 14 dimensional 5-brane
theory is consistent and exists from another different {\em route} : 
Assuming that S duality is an exact symmetry of the super
p-branes then 
we can derive some important consequences about their
quantization,
at D dimensions, a (d-1) brane is dual to another (\~{d}-1)
brane \cite{r11}, for
\begin{equation}
{\tilde d} = D -d - 2,
\end{equation}
whatever this condition seems trivial, we require
for its validity the existence , a priori, of both the (d-1)  
and the (\~{d}-1) brane then we have only the following cases :
\begin{eqnarray}
D=14~~~~~~~d=6~\mbox{and}~{\tilde d}=6~.\\
D=10~~~~~~~d=2~\mbox{and}~{\tilde d}=6~.\\
D=6~~~~~~~~d=2~\mbox{and}~{\tilde d}=2~.
\end{eqnarray}
We require the existence of  both the d and the {\~{d} theory 
 because S duality maps a (d-1) brane
at weak coupling to a ({\~{d}-1) brane at strong coupling and
vice-versa. So, if one of these dual theories does not exist
at this specific D, then our theory will be anomalous.
Actually, this condition was too effective to rule out 
all the p branes except p=2 and 5 only.
It indicates that superstring theory 
is consistent in 6 and 10 dimensions \cite{r14} 
whereas the super 5 brane is allowed only for 10 and 14 dimensions.
We notice that these dimensions are exactly the ones where the gravitational
anomaly exists (D=4k+2 for integer k).
It remains to confirm this result on  a stiff ground by
an explicit anomaly calculation.

\vspace{3cm}

I am very grateful, especially, to Prof. P. Rotelli, 
and generally, to all the members
of the physics department at Lecce university for their kind 
hospitality.

\newpage
\begin{table}
\caption{The complete classification of real Clifford 
algebra in D dimensions 
with signature (s,t) where the real number of components
 (comp) $<$ 64.
\label{table1}}
\begin{tabular}{l|ccccccccccccr}
s$/$t\tablenote{($^2$A) means A$\oplus$A}
&0&1&2&3&4&5&6&7&8&9&10&11&12\\
\tableline
0&$\pm$1&R&C&H&$^2$H&H(2)&C(4)&R(8)&$^2$R(8)
&R(16)&C(16)&X&X\\
1&R&$^2$R&R(2)&C(2)&H(2)&$^2$H(2)&H(4)
&C(8)&R(16)&$^2$R(16)&R(32)&X&X\\
2&C&R(2)&$^2$R(2)&R(4)&C(4)&H(4)&$^2$H(4)&
H(8)&C(16)&R(32)&$^2$R(32)&X&X\\
3&H&C(2)&R(4)&$^2$R(4)&R(8)&C(8)&H(8)
&$^2$H(8)&X&X&X&X&X\\
4&$^2$H&H(2)&C(4)&R(8)&$^2$R(8)
&R(16)&C(16)&X&X&X&X&X&X\\
5&H(2)&$^2$H(2)&H(4)&C(8)&R(16)
&$^2$R(16)&R(32)&X&X&X&X&X&X\\
6&C(4)&H(4)&$^2$H(4)&H(8)&C(16)&R(32)
&$^2$R(32)&X&X&X&X&X&X\\
7&R(8)&C(8)&H(8)&$^2$H(8)&X&X
&X&X&X&X&X&X&X\\
8&$^2$R(8)&R(16)&C(16)&X&X&X&X&X&X
&X&X&X&X
\\
9&R(16)&$^2$R(16)&R(32)&X&X&X&X&X&X&X&X&X&X\\
10&C(16)&R(32)&$^2$R(32)&X&X&X&X&X&X&X&X&X&X\\
11&X&X&X&X&X&X&X&X&X&X&X&X&X\\
12&X&X&X&X&X&X&X&X&X&X&X&X&X\\
\end{tabular}
\end{table}
\newpage
\begin{table}
\caption{The allowed minimal spinor (M) 
for D dimensions 
with signature (s,t) where the real number of components $<$ 64. 
We present also
the possible number of extended supersymmetry N and the allowed 
p-branes ($p=d-1$) with their extended supersymmetry n.
\label{table2}}
\begin{tabular}{l|ccccc}
D&(s,t) 
\tablenote{s for space and t for time, obviously, the (t,s) 
case is also allowed.}& 
M\tablenote{real components.}&The automorphism group
&N\tablenote{again, we will restrict 
ourselves to the only physical cases i.e N$\leq$8.}&(p,n)\\
\tableline
1&(1,0)&1&SO(1)&1...8\\
2&(1,1)&1&SO(1)$_+\times$SO(1)$_-$&1...8\\
3&(2,1)&2&SO(2)&1...8&(1,2)\\
4&(2,2)&2&SO(2)$_+\times$SO(2)$_-$&1...8&(1,4),~(2,1)\\
5&(3,2)&4&SO(4)&1...8&(2,2),~(3,1)\\
6&(3,3)&4&SO(4)$_+\times$SO(4)$_-$&1...8&(1,8),~(3,2)\\
7&(7,0), (4,3)&8&SO(8)&1...4&(2,4),~(4,1)\\
8&(8,0), (4,4)&8&SO(8)$_+\times$SO(8)$_-$&1...4&(3,4),~(5,1)\\
9&(9,0), (8,1), (5,4)&16&SO(16)&1, 2&(4,2)\\
10&(9,1), (5,5)&16&SO(16)$_+\times$SO(16)$_-$&1, 2&(1,16),~(5,2)\\
11&(10,1), (9,2), (6,5)&32&SO(32)&1&(2,8)\\
12&(10,2), (6,6)&32&SO(32)$_+\times$SO(32)$_-$&1&(3,8)\\
\end{tabular}
\end{table}


\begin{references}

\bibitem{r1} J.~A.~Harvey, ``Magnetic Monopoles, Duality, 
and Supersymmetry'', hep-th/9603086.

\bibitem{r2} J.~H.~Schwarz, ``The Power of M Theory'', 
hep-th/9510086. 

\bibitem{r3} C.~Vafa, ``Evidence of F Theory'', 
hep-th/9602022.
\\ E.~Witten, ``Phase Transitions in M-Theory and 
F-Theory'', hep-th/9603150.

\bibitem{r7} J.~Strathdee, Int. J. of Mod. Phys. {\bf 2A} 
(1987) 273.

\bibitem{r4} I.~R.~Porteous, Clifford Algebras and the 
Classical Groups, Cambridge University Press, 1995.

\bibitem{r5} I.~M.~Benn~and~R.~W.~Tucker, An Introduction to 
Spinors and Geometry with Applications in Physics, Adam 
Hilger, 1987.

\bibitem{r6} T.~Kugo~and~P.~Townsend, Nucl. Phys. 
{\bf B221} (1983) 357.

\bibitem{r8} I.~Bars, ``Duality and Hidden dimensions'', 
hep-th/9604200.

\bibitem{r9} S.~V.~Ketov,~H.~Nishino~and~S.~J.~Gates,~Jr., 
Nucl. Phys. {\bf B393} (1993) 149, hep-th/9207042.

\bibitem{r10} E.~Sezgin, ``Self-Duality and Supersymmetry'', 
hep-th/9212092.

\bibitem{r11} M.~J.~Duff,~R.~R.~Khuri,~J.~X.~Lu, 
Phys.~Rep. {\bf 259} (1995) 213.

\bibitem{r12} E.~Bergshoeff,~E.~Sezgin,~P.~K.~Townsend, 
Phys. Lett.~B189 (1987) 75.

\bibitem{r13} M.~J.~Duff~and~J.~X.~Lu, 
Nucl. Phys. {\bf B390} (1993) 276, hep-th/9207060. 
\bibitem{r14} J.~H.~Schwarz, ``Anomaly-Free Supersymmetric
Models in Six Dimensions'', hep-th/9512053.
\end{references}
\end{document}